\documentclass[grl]{AGUTeX}

\usepackage{url}
\usepackage{graphicx}
\usepackage{subfigure}

\newcommand{\dgr}{{$^\circ$}}

\setkeys{Gin}{draft=false}
\authorrunninghead{ }
\titlerunninghead{published in Geophysical Research Letters, 40, 2913--2917, doi:10.1002/grl.50617, 2013}

\begin{document}

%
%

\title{Proton entry into the near-lunar plasma wake for magnetic field aligned flow}
%
%

%
%



 \authors{M. B. Dhanya,\altaffilmark{1} A. Bhardwaj,\altaffilmark{1}
  Y. Futaana,\altaffilmark{2} S. Fatemi,\altaffilmark{2,4} M. Holmstr\"{o}m,\altaffilmark{2} S.~Barabash,\altaffilmark{2}
  M. Wieser,\altaffilmark{2} P. Wurz,\altaffilmark{3} A. Alok,\altaffilmark{1} and R. S. Thampi \altaffilmark{1}}
 
 \altaffiltext{1}{Space Physics Laboratory, Vikram Sarabhai Space Center, Trivandrum, India.}
 \altaffiltext{2}{Swedish Institute of Space Physics, Box 812, Kiruna SE-98128, Sweden.}
 \altaffiltext{3}{Physikalisches Institut, University of Bern, Sidlerstrasse 5, CH-3012 Bern, Switzerland.}
  \altaffiltext{4}{Department of Computer Science, Electrical and Space Engineering, Lule\aa{} University of Technology, Lule\aa{}, Sweden}
 






%
%


\begin{abstract}
We report the first observation of protons in the near (100--200 km from surface) and deeper (near anti-subsolar point) lunar plasma wake  when the interplanetary magnetic field (IMF) and solar wind velocity ($v_{\textup{\small{sw}}}$) are parallel (aligned flow; angle between IMF and $v_{\textup{\small{sw}}}$ $\le$ 10\dgr ). More than 98\% of the observations during aligned flow condition showed the presence of protons in the wake. These observations are obtained by the SWIM sensor of the SARA experiment on Chandrayaan-1.  The observation cannot be explained by the conventional fluid models for aligned flow. Back-tracing of the observed protons suggests that their source is the solar wind. The larger gyro-radii of the wake protons compared to that of solar wind suggest that they were part of the tail of the solar wind velocity distribution function. Such protons could enter the wake due to their large gyro-radii even when the flow is aligned to IMF.  However, the wake boundary electric field may also play a role in the entry of the protons in to the wake.

\end{abstract}

%
%

%

\begin{article}

%
%
 \section{Introduction}
 
 Characteristics of lunar plasma wake at distances in the range of a fraction of lunar radius (near-wake) to few lunar radii (far-wake) is an evolving problem. The limited observations and simulations performed for far-wake have shown that the geometry of the wake depends on the orientation of interplanetary magnetic field (IMF)
  \cite{Ogilvie96,Birch02,Travnicek05,Kallio05,Wiehle11,Fatemi13}. 
 In the near-wake, protons have been observed to enter the wake along IMF  \citep{Futaana10b},  as well as perpendicular to IMF \citep{Nishino09a,Nishino09b,Wang10}, when the IMF is predominantly perpendicular to the solar wind velocity. The entry perpendicular to IMF requires either the convective electric field of solar wind \citep{Nishino09b,Wang10} or an ambi-polar potential drop across the wake boundary \citep{Nishino09a}  to assist the entry of protons in to the wake. The situation when solar wind velocity vector is aligned to the IMF (aligned flow) is a special case where none of the above mechanisms can transport protons to deeper locations inside the near-lunar wake. Recent \citep{Holmstrom12} and past simulations \citep{Michel68,Wolf68,Spreiter70}  for the far-lunar wake have shown that the entry of solar wind protons to the near-wake will be inhibited during aligned flow. However, one can expect solar wind proton entry due to their own gyro motion (provided gyro-radius is larger) even under parallel IMF, but no observations have been reported. We report the unique events under weak IMF and high proton temperature that make gyro-radius larger and thus enable solar wind protons to enter the deeper wake. These first observations during aligned flow have been made by the SWIM/SARA experiment aboard Chandrayaan-1.
    
  \section{Instrument and Data}
  
\begin{figure}
     		   \centering
  		 		     \includegraphics[width=20pc]{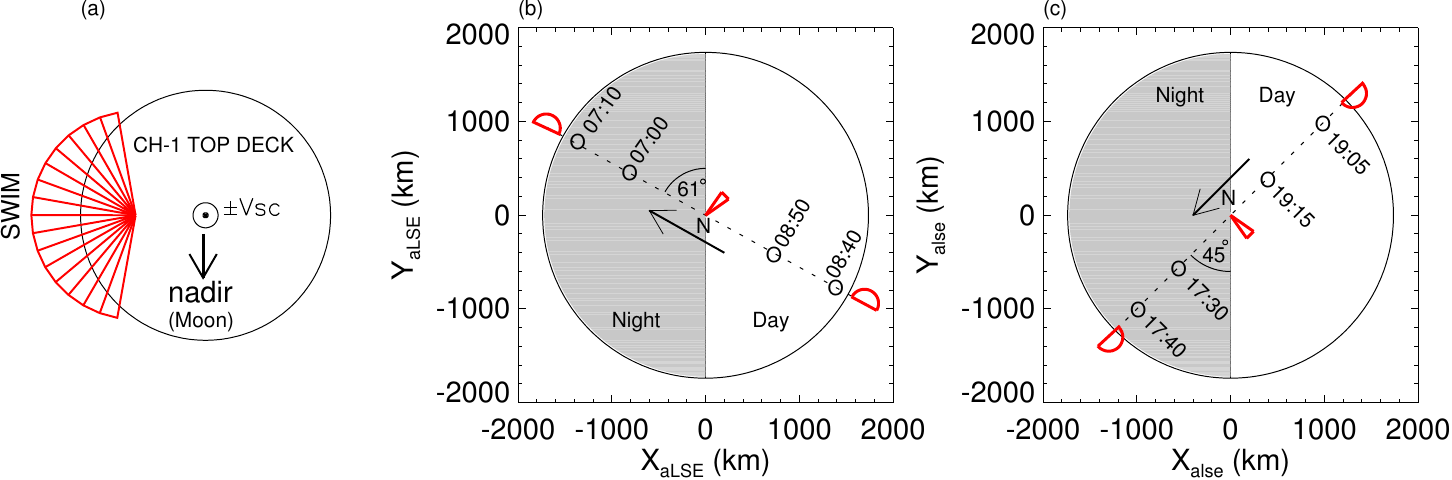}
     		 		      \caption{(a) Field of view (FoV) of SWIM on Chandrayaan-1 (CH-1). The direction of nadir and velocity vector of spacecraft ($v_{\textup{\small{sc}}}$) are shown. (b) SWIM FoV orientation (shown in red) as seen from the lunar north pole (indicated as `N') for an orbit (dotted line) on 18 July 2009 between 07:00 UT and 08:50 UT. The angle between orbital plane of CH-1 and the day-night terminator plane (Sun-aspect angle) was 61\dgr. SWIM FoV plane is perpendicular to the ecliptic plane at the poles and is in the ecliptic plane closer to the equator. The nightside is shown by grey shaded area. The arrow indicates the direction of motion of CH-1. (c) Same as (b) for 30 April 2009 (17:30--19:15 UT) when the Sun-aspect angle was 45\dgr. }
     		 		        \label{fig:swimfov}
  \end{figure}

  The SWIM (Solar WInd Monitor) was an ion-mass analyser and was one of the two sensors of the SARA (Sub-keV Atom Reflecting Analyser) experiment on the 100--200 km polar orbiting lunar spacecraft Chandrayaan-1 \citep{Bhardwaj05, Barabash09, McCann07}.
  SWIM has a fan shaped field of view (FoV) having 16 angular pixels with an angular resolution around 10\dgr $\times$ 4.5\dgr\ (see Fig.~\ref{fig:swimfov}a for illustration). The FoV orientation is illustrated in the Figs.~\ref{fig:swimfov}b and \ref{fig:swimfov}c for two orbits when the orbital plane makes an angle of 61\dgr\ and 45\dgr\, respectively, with the day-night terminator plane.
  
   SWIM 
was operated in the energy range 100--3000 eV with an energy resolution of $\Delta E/ E \sim$7\%. The SWIM observations provide differential flux of ions in a specific energy and angular (direction) bin every 31.25 ms.  The time taken for a full angular and energy scan was 8 s. Only the observations made when the Moon was located  in the upstream solar wind (outside Earth's bow shock) have  been considered in the present analysis.  The co-ordinate system used for the analysis is `aberrated LSE co-ordinates' (aLSE), where the \emph{x}-axis is along the anti-solar wind velocity direction, the \emph{z}-axis is normal to the solar wind velocity vector towards the ecliptic north, and the \emph{y}-axis completes the right handed co-ordinate system. For the upstream solar wind parameters, such as solar wind velocity, density, and IMF orientation, we have used level-2 data from the SWEPAM and MAG instruments on the ACE satellite. Since ACE makes measurements near L1 point, the data have been time shifted by considering the solar wind speed 
and distance of ACE from the Moon. For the analysis in this paper, we considered the flow to be aligned when the angle between the solar wind velocity and IMF was within $\pm$10\dgr. 
  
  \section{Observations}

\begin{figure}
  			  		   \centering
  				     \includegraphics[width=23pc]{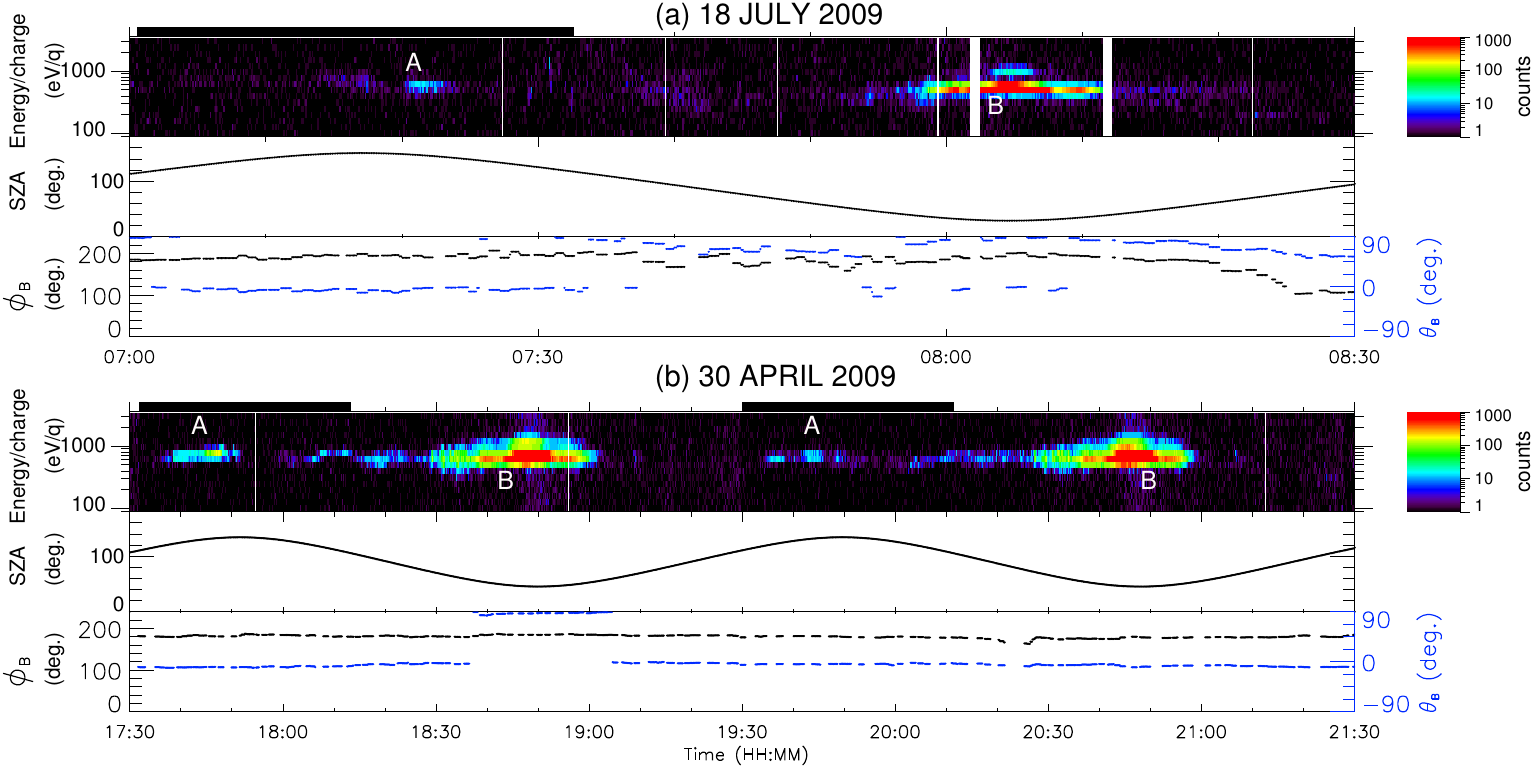}
  			  		 		      \caption{(a) SWIM Observation on 18 July 2009 when the orbit of Chandrayaan-1 (CH-1) was at an altitude of 200 km. The top panel is the energy-time spectrogram of proton counts with the black boxes overlaid representing the time interval when CH-1 was in the lunar plasma wake. The population marked as A represents protons observed in the wake and B represents upstream solar wind protons. The middle panel represents the solar zenith angle (SZA), and the bottom panel represents the  orientation of IMF in terms of azimuth and elevation angles. The azimuth angle $\phi_B$ is the angle between projection of IMF in the \emph{x-y} plane and the \emph{x}-axis, whereas the elevation angle  $\theta_B $ is the angle between the \emph{z} component of IMF and the \emph{x-y} plane, so that +90\dgr\ refers to orientation towards north and $-$90\dgr\ towards south (all in aLSE frame). The orbit of CH-1 for this event is shown in Fig.~\ref{fig:swimfov}b (b) Same as (a) for 30 April 2009 when CH-1 was at an altitude of 100 km. The orbit of CH-1 for this event is shown in Fig.~\ref{fig:swimfov}c}
  			  		 	   \label{fig:ETspectra}		  	 		   		     			  \end{figure}

  			The energy-time spectrogram of the proton counts observed by SWIM at 200 km altitude above the lunar surface on 18 July 2009,  when the orbital plane of Chandrayaan-1 was at an angle of 61$^\circ$ from the day-night terminator,  is shown in  Fig.~\ref{fig:ETspectra}a.  The two populations of protons observed in this orbit are marked as A and B in the Fig.~\ref{fig:ETspectra}a. Our emphasis here is on the population A observed in the wake when the solar zenith angle (SZA) was close to 140\dgr. The orientation of IMF (Fig.~\ref{fig:ETspectra}a, bottom panel) was aligned with solar wind velocity within $\pm$5$^\circ$. The configuration of SWIM field of view during this orbit is shown in Fig.~\ref{fig:swimfov}b. Hereafter, we refer to this observation as Event-1. The population B corresponds to upstream solar wind protons observed on the lunar dayside.  			
  			
  	        Two events observed on 30 April 2009, in two consecutive orbits, are shown in Fig.~\ref{fig:ETspectra}b, when the orbital plane  of Chandrayaan-1 was at an angle of 45\dgr \ from the day-night terminator and at an altitude of 100 km. For the orbit which spans 17:20--19:20 UT, intense proton counts were observed in the wake around 17:35--17:55 UT (indicated as A) when the SZA  was close to 130\dgr.\ The IMF was within $ \pm $10\dgr \ of the solar wind
  	        velocity vector.  We refer to this observation as Event-2. Similarly, in the following orbit, proton counts are observed during 19:35--19:55 UT in the wake (indicated as A), when the IMF and solar wind velocity were aligned (within $ \pm5$\dgr). Fig.~\ref{fig:swimfov}c shows the FoV orientation of SWIM for these orbits. We refer to this observation as Event-3.  The summary of the three events along with the upstream solar wind parameters are given in Table 1.

  	        Over the six months (January--July 2009) of SWIM observations, the condition for the aligned flow (the angle between the solar wind and IMF is less than 10\dgr) was satisfied on 30 days in 66 orbits when SWIM was in the wake. More than 98\% of the observations  in the lunar wake registered non-zero counts when the aligned flow condition was satisfied. In the analysis, counts=1 (in every energy and direction bin) was considered as background and was subtracted from the observed counts.  
  	       			 
  		\section{Discussion}
  		
  		For a perfect magnetic field aligned flow the convective electric field ({\bf E}$_{conv}$) is zero. For an angle of $\le$10\dgr \ between IMF and solar wind velocity the convective electric field would be negligible. Hence, the transport of  solar wind protons scattered on the dayside to the nightside under the influence of {\bf E}$_{conv} \times ${\bf B}$_{\rm IMF}$ (gyro-radius larger than the Moon radius, e.g., \cite{Nishino09b}) would not be efficient for an aligned flow. Since the solar wind velocity and IMF are parallel, the entry of protons parallel to IMF (e.g., \cite{Futaana10b})  also cannot explain the reported proton observations in the deep near-lunar wake. The fluid approximation for aligned flow \citep{Michel68,Wolf68}, as well as the continuum MHD approach \citep{Spreiter70} can not explain the observed protons because they are observed in the near and deeper wake. Classical theory \citep{Spreiter70} and the recent hybrid simulations \citep{Holmstrom12} show that the entry of solar wind protons is inhibited when the IMF is aligned with the solar wind flow resulting in an elongated wake. However, models that emphasise on the far-wake have limited applicability in the near-wake because  in the near-wake the plasma dynamics is much more complicated due to the presence of processes such as lunar surface charging \citep{Halekas11a}.
  		  			
\begin{figure}
  	     	\centering
  	  	 	\includegraphics[width=20pc]{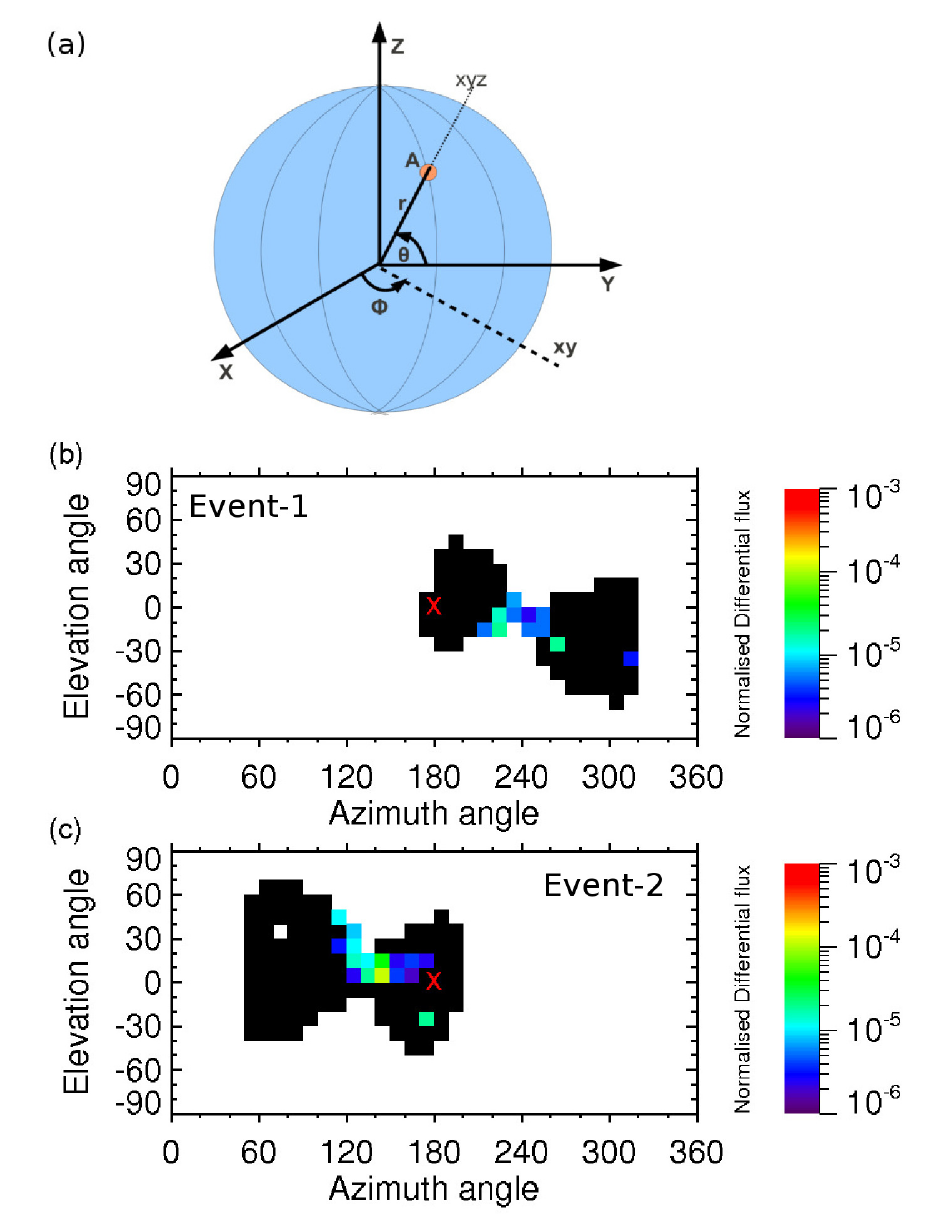}    		   		  	 	 
  	 	 \caption{(a) The spherical polar co-ordinate system (origin at the centre of Moon) used to represent the velocity distribution of the wake protons, where \emph{$\phi$} is the velocity vector azimuth angle [0\dgr, 360\dgr]  in the \emph{x-y} plane, and \emph{$\theta$} is the velocity elevation angle [$-$90\dgr, $+$90\dgr]  as measured from \emph{z-axis} of aLSE. The position `A' represents an arbitrary location of SWIM/Chandrayaan-1. (b) Velocity distribution of protons for Event-1, where the colour bar represents the differential flux normalised to the solar wind flux (normalised differential flux). The black filled areas represents zero counts observed by SWIM. The direction of solar wind (from ACE) is shown by the cross (red) symbol. (c) Same as (b) for Event-2.}
 	   		  	 	
  	  		   		  	 	  \label{fig:vel_distr}  	    		   		  	 	    \end{figure}

  	  	The observed energy provides the speed of the protons and the viewing direction provide the direction of travel of the particle in the SWIM instrument frame. We have computed the velocity distribution of the observed protons in the aLSE frame, using  full wake observations. For this we have defined a spherical coordinate system (\emph{v, $\phi$, $\theta$}) in the velocity space (see Fig.~\ref{fig:vel_distr}a for illustration), where \emph{v} is the velocity magnitude, \emph{$\phi$} is the velocity vector azimuth angle ([0, 360] degrees), and \emph{$\theta$} is the velocity elevation angle ([$-$90, $+$90] degrees).
  	  
  	  	 The velocity distribution for Event-1 and Event-2 are shown in Figs.~\ref{fig:vel_distr}b and 3c, respectively. 
  	  	  It can be seen from Fig.~\ref{fig:vel_distr}b that for the Event-1, most of the protons observed by SWIM were travelling in the direction of $\phi$: 210--260\dgr\ and $\theta$: +20 to $-$20\dgr. The energy is around 700--800 eV. For Event-2, which happened on the dawn hemisphere (whereas Event-1 was observed in the dusk side hemisphere), SWIM has observed protons travelling mostly in the direction of $\phi$: 120--180\dgr\ and $\theta$: 0\dgr\ to +40\dgr. Also for this event most of the observed  protons have energy around 700 eV. According to the ACE data, the average energy of the solar wind was $\sim$420 eV for Event-1 (for an average solar wind speed of 290 km s$^{-1}$, cf. Table 1) and $\sim$560 eV for Event-2 (for an  average solar wind speed of 335 km s$^{-1}$, cf. Table 1). The angular velocity distribution of the protons for Event-3 was similar to that of the Event-2, and therefore is not shown here.
  	  	 
  	  	 To understand the origin of the nightside ions, we back-traced the observed protons from the location of their detection using the observed velocity magnitude and direction. A similar method had been used, for example, by \citep{Futaana03}, for studying Moon-solar wind proton interaction. Here the back-tracing was carried out with uniform (constant in time and position) upstream parameters as shown in Table 1. We assumed, for simplicity, that the magnetic field direction is perfectly parallel to the solar wind velocity vector.

\begin{figure}
  	  	   	  	   	    	 \centering
  	  	   	  			 \includegraphics[width=23pc]{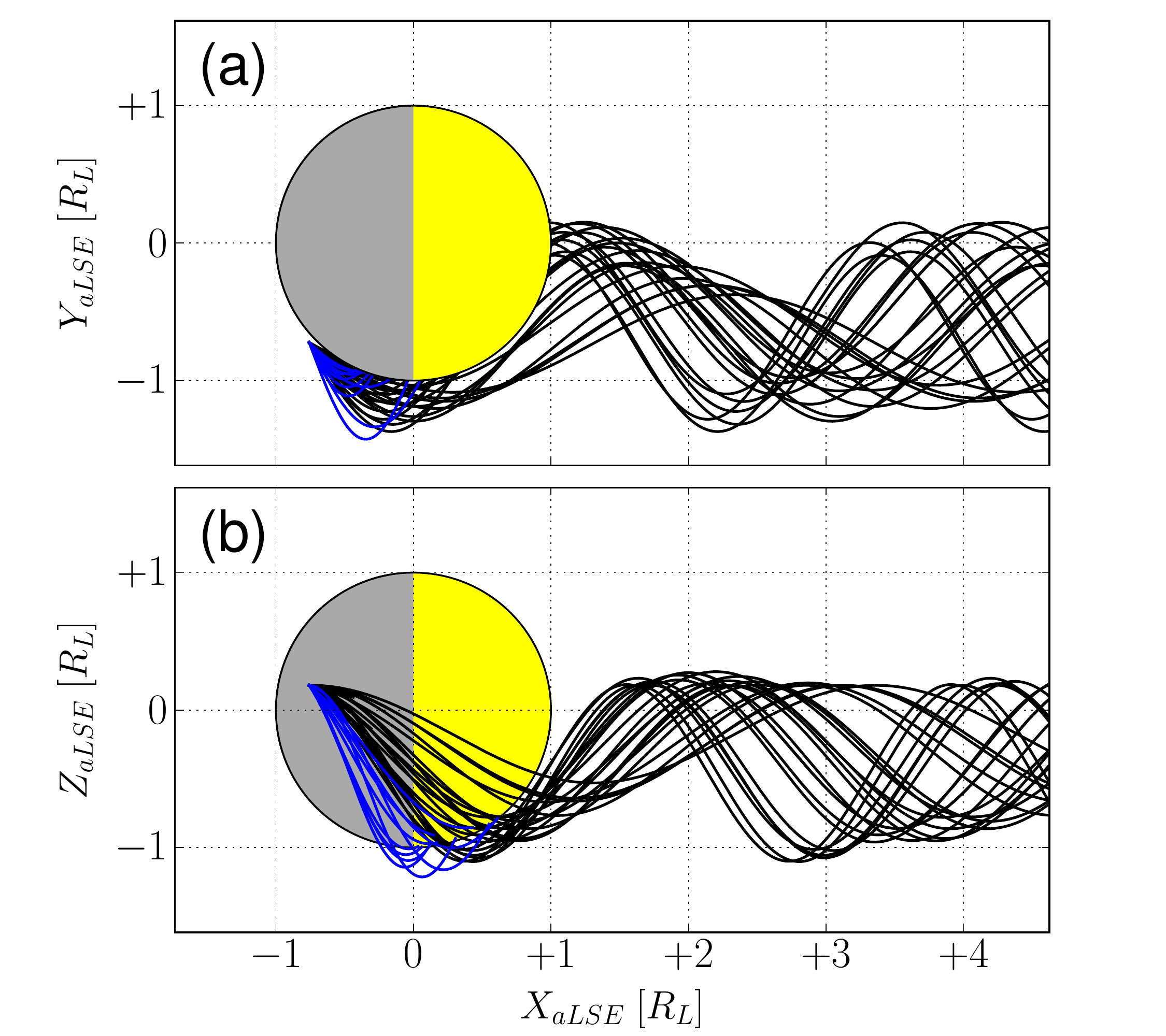}
  	  	   	  	   	  					 	\caption{(a) Back-traced trajectories of observed protons in Event-2 projected onto the $x$-$y$ plane. The black curves show trajectories originating in the solar wind, and the blue curves show trajectories originating from the lunar surface. The unit of horizontal and vertical axes are in lunar radii ($R_{L}$) (b) Same as (a) with trajectories projected onto the $x$-$z$ plane  }
  	  	   	  	   	    	 				\label{fig:trajectory}
  	  	   	  	   	    	 \end{figure}

  	  	  As an example, Fig.~\ref{fig:trajectory} shows projections in two planes ($x$-$y$ and $x$-$z$) of the back-traced trajectories of the observed protons during Event-2. These trajectories show that the majority of observed protons could have originated from the solar wind. Some particles may come from the lunar surface, but the fraction is small. These solar wind ions could enter the wake due to their larger gyro-radii even when the flow is aligned to IMF. The larger gyro-radii of the wake protons compared with the solar wind gyro-radii ($\sim$95 km in Event- 2 case) suggest that they were part of the tail of the solar wind velocity distribution function. Only uniform electric and magnetic fields are assumed in the entire simulation domain. However, we examined the back-tracing using an enhanced IMF magnitude (by a factor of 1.5 as an upper limit) to simulate the enhancement of the magnetic field in the wake \citep{Colburn67}. The back-tracing results were almost identical. We also examined small angles between solar wind velocity and IMF in the range 5\dgr\ to 10\dgr, and the results did not change significantly as well. This backtracing shows that the wake ions are likely of the solar wind origin. On the other hand, how much the solar wind can penetrate into the deep near-wake is a further study. For that, the comparison with hybrid simulation would be a good tool (e.g. \citep{Fatemi12}).
  	  	It should be noted that an ambipolar electric field formed at the boundary of the lunar wake \citep{Ogilvie96,Futaana01}  pointing inward can help the entry of solar wind protons in to the wake. Thus, the observed cross-field entry of the solar wind protons may be a manifestation of the combination of thermal motion and wake boundary electric field. 
  	  	 	    	
	      The presence of these protons can alter the electrodynamics in the near-wake region. The lunar surface charging on the nightside may also be affected due to these protons, which otherwise is negatively charged to potentials of around few hundred volts \citep{Halekas11a}. These protons, possibly from the high energy tail of the solar wind, can cause sputtering if they hit the lunar nightside surface. According to \citep{Wurz07} the total sputter yield per incident solar wind ion for the lunar surface is typically in the range 2$\times$10$^{-5}$ to 0.07. For an incident solar wind flux of $\sim$10$^8$ cm$^{-2}$ s$^{-1}$ the sputtered flux is around $\sim$10$^7$ cm$^{-2}$ s$^{-1}$ corresponding to a total contribution to the exospheric density of up to 10 cm$^{-3}$ \citep{Wurz07}. Since in our observation the protons are observed mostly in three viewing directions of SWIM, assuming an angular extend of the distribution of 30\dgr$\times$30\dgr, a crude estimate of the flux of the protons for Event-1 is $\sim$.1$\times$10$^6$ cm$^{-2}$ s$^{-1}$. Assuming that all of them hit the nightside surface, a crude estimate of the upper limit for contribution to the exospheric density is $\sim$0.1 cm$^{-3}$. For Event-2 and Event-3 the contribution will be  higher due to the higher differential flux of observed wake protons. This estimate, although crude, indicates that the sputtered flux and thus the contribution to the exosphere, will be very low. A detailed and more accurate estimation can be addressed in a future work. These results applies not only to the Moon, but also to similar atmosphereless  planetary body such as some of the satellites of the outer planets.
	      
	      \begin{table}
	      \caption{{Summary of the Events and the upstream solar wind parameters}}
	      \centering
	      \begin{tabular}{lccccccc}
	      \hline
	      Event  & SAA(\dgr) & Time (UT) & SZA (\dgr) & V$_{sw}$ & n$_p$  & T$_{p}$ &  IMF \\
	      \hline
	       Event-1 & 61 & 07:20--07:25 & 160--140 & 290 & 3 & 2$\times$10$^4$ & 1  \\
	       Event-2 & 45 & 17:35--17:55 & 120--140 & 335 & 6 & 4.5$\times$10$^4$ & 3  \\
	       Event-3 & 45 & 19:35--19:55 & 120--140 & 335 & 6 & 4.5$\times$10$^4$ & 3  \\
	      \hline
	      \end{tabular}
	     \tablenotetext{}{SAA= Sun aspect angle, which is the angle between the orbital plane of Chandrayaan-1 and the day-night terminator plane, SZA= solar zenith angle, V$_{sw}$= solar wind speed in units of km s$^{-1}$ (ACE), n$_{p}$= solar wind proton density in units of cm$^{-3}$ (ACE), T$_{p}$= solar wind proton temperature in units of \dgr K (WIND), IMF= Interplanetary Magnetic Field in units of nT (ACE).}
	      \end{table}

  	\section{Conclusion}
  	
       We have reported observations of protons in the near and deeper lunar wake during the aligned flow. These observations cannot be explained by the conventional fluid models for aligned flow. Back-tracing of the observed protons suggests that their origin could be the solar wind. The computed larger gyro-radii of the wake protons compared to that of solar wind implies that these must be from the tail  of the solar wind velocity distribution. Such protons could indeed enter the wake even during aligned flow  due to their larger gyro-radius. However, the wake boundary electric field may assist the entry of the protons into the wake. These protons can alter the electrodynamics in the near-wake region and may contribute towards the nightside exosphere by causing sputtering.


%
%
%
%
%
%
%

\begin{acknowledgments}
	The ACE data is provided by the ACE Science Center. The efforts at Space Physics Laboratory of Vikram Sarabhai Space Centre are supported by Indian Space Research Organisation (ISRO). The efforts at the Swedish Institute of Space Physics were supported in part by European Space Agency (ESA) and National Graduate School of Space Technology (NGSST), Lule\aa, Sweden, and the Swedish Research Links Programme funded by the Swedish International Developement Corporation Agency (SIDA). The effort at University Bern is supported in part by ESA and by the Swiss National Science Foundation. 
\end{acknowledgments}

\end{article}

\end{document}